\documentclass[runningheads]{llncs}
\usepackage[left=4cm, right=4cm,top=4cm]{geometry}
\title{thesis}
\usepackage{xcolor}
\usepackage{graphicx}
\usepackage{epstopdf}
\usepackage{subfig}
\usepackage{algorithm}
\usepackage{algorithmic}
\usepackage{amsmath,amssymb}
\usepackage{geometry}
\usepackage{CJK}
\usepackage{hyperref}
\usepackage{url}
\usepackage{float}
\usepackage[english]{babel}
\hypersetup{
    citebordercolor=blue,
    linkbordercolor=orange
}

\usepackage{dsfont}
\usepackage{caption}
\usepackage{multirow}
\usepackage{dsfont}
\usepackage{setspace}
\usepackage{url}
\usepackage{braket}

\usepackage{graphicx}

\begin{document}
\title{The item selection problem for user cold-start recommendation}
%
%
\author{Yitong Meng \and
Jie Liu\and Xiao Yan\and James Cheng}

\institute{The Chinese University of Hong Kong\\
mengyitongge@163.com}

\maketitle              

When a new user just signs up on a website, we usually have no information about him/her, i.e. no interaction with items, no user profile and no social links with other users. Under such circumstances, we still expect our recommender systems could attract the users at the first time so that the users decide to stay on the website and become active users. This problem falls into new user cold-start category and it is
crucial to the development and even survival of a company.

Existing works on user cold-start recommendation either require additional user efforts, e.g. setting up an interview process~\cite{DBLP:conf/iui/RashidACLMKR02}, or make use of side information~\cite{DBLP:conf/recsys/SedhainSBXC14} such as user demographics, locations, social relations, etc. 
However, users may not be willing to take the interview and side information on cold-start users is usually not available. 
Therefore, we consider a pure cold-start scenario where neither interaction nor side information is available and no user effort is required.
Studying this setting is also important for the initialization of other cold-start solutions, such as initializing the first few questions of an interview.

\section{Problem formulation}
Our model is built upon the output of latent factor models~\cite{PMF,he2017neural,liang2018variational}.
Latent factor model is a powerful tool that can embed the user and item information in a low-dimension space. A user's preference on an item is propotional to the inner product of their latent vectors. We assume the latent vectors of warm users and items are given by some state-of-the-art latent factor model. 
Our problem setting is rather simple: given the latent vectors of warm users $\mathcal{U}$ and the latent vectors of all items $\mathcal{X}$, our goal is to predict a small item set $\mathcal{Y}\subset \mathcal{X}$ which is most likely to contain an favorite item of a future user. Our goal is to identify the favorite items because according to our daily experience, a first-timer would continue to explore a website if he could find one of his favorite in the very beginning. The recommended item set should also be small because we want a user could review it conveniently.

The main challenge is that we have no information about the cold-start users, then how should we estimate their favorite?
We detour this problem by aggregating the favorite items of warm users, and expect that the aggregated item set contains an item that a cold-start user favorite. The underlying assumption is that the cold-start users and warm users come from the same distribution.
We formulate the problem as follows.

\medskip
 \textbf{Problem formulation}. Given the latent vectors of a set of $W$ warm users $\mathcal{U} = \{u_w \in \mathbb{R}^D\}_{w=1}^W$, a set of $N$ items $\mathcal{X} = \{x_n\in \mathbb{R}^D\}_{n=1}^N$ , we aim to find a subset $\mathcal{Y} \subset  \mathcal{X}$ with size $M$, such that,
\begin{align}\label{eq:item_selection}
    \mathcal{Y}^{*}=\operatorname{argmin}_{\mathcal{Y}:\{|\mathcal{Y}| = M, \mathcal{Y} \subseteq \mathcal{X}\}} fav\_loss(\mathcal{Y}),
\end{align}
where,
\begin{align}\label{eq:favloss}
    fav\_loss(\mathcal{Y})=\sum_{u\in \mathcal{U}} l(u, \mathcal{Y})
\end{align}
and 
\begin{align}\label{eq:l}
    l(u, \mathcal{Y}) = \max_{x \in \mathcal{X}} u^\mathsf{T}x - \max_{y \in \mathcal{Y}} u^\mathsf{T}y.
\end{align}
Eq \ref{eq:l} evaluates the difference between a user's favorite item in $\mathcal{X}$ and his favorite item in $\mathcal{Y}$, and we want the difference as small as possible. Our problem definition can be deemed as an extension of traditional \textit{Maximum Inner Product Search (MIPS)} problem~\cite{shrivastava:alsh}. Tradition MIPS  considers each query separately while our problem considers a set of queries $\mathcal{U}$ simultaneously.

When $M\geq W$, there is a trivial solution by picking up each user's favorite item in $\mathcal{X}$. However, the normal case is $M\ll W$ and we need to explore $C_{N}^M$ possibilities, which is computationally expensive.
For such case, we propose several possible solutions, including \textit{Max-Norm}, \textit{Max-In-Degree}, \textit{User-Expectation}, \textit{Inner Product Graph Search (IPGS) }, \textit{Submodular Greedy Algorithm (Submodular)} and \textit{Convex Hull Approximation}. 
The Submodular and the convex hull approximation methods are principled methods that directly solve Eq \ref{eq:item_selection}, while the other four methods are heuristic.
The Submodular method employs a greedy algorithm to approximate the optimal solution, based on the fact that $fav\_loss(\mathcal{Y})$ is a monotonic submodular function.
The convex hull approximation method try to find the convex hull of $\mathcal{X}$, which is the optimal solution if the cardinality of the convex hull is less than $M$.
Max-Norm selects the items with the largest vector norm, which is inspired by the well-known \textit{norm-bias} phenomenon in the MIPS problem~\cite{liu:understanding}.
Max-In-Degree tends to select the items that are similar to many other items.
The User-Expectation method predicts a cold-start user's latent vector by averaging the latent vectors of all warm users and then use the mean vector to query items.
IPGS statistics the frequency of the favorite items of all warm users, and recommends the most frequent ones.
We describe the six methods in the following section.
\section{Solutions to the proposed problem}

\subsection{Max-Norm method}
\medskip
Eq \ref{eq:item_selection} can be considered as a extension of the MIPS problem by considering a group of queries at the same time.
It is well-known that MIPS problem is biased towards large-norm items. To illustrate, consider the computation of inner product between user $u$ and item $x$,
\[
u^{\mathsf{T}}x = \Vert u \Vert \Vert x \Vert \cos{\alpha}.
\]
There are two factors affecting the final value of inner product, i.e. the vector norms and the angle between them. 
Norm-bias~\cite{liu:understanding} in MIPS problem means that large norm items are much more likely to be the results of MIPS.
The paper of Yan and Liu, et al.~\cite{liu:understanding} shows items ranking top 5\% in norm take up nearly 90\% in the ground truth top-10 MIPS results for Yahoo!Music and WordVector dataset. They also show that the norm bias is caused by skewed norm distribution, in which the top ranking items have much larger norms than the others. 

Considering the norm-bias in our problem, 
we plot the relationship between the norm of item vectors and the number of the high ratings that each item receives on three recommendation datasets, MovieLens-20M, Epinions and Amazon's All Beauty, whose detailed descriptions can be found in section \ref{sec:usercold_datasets}. The y-axis is the norm, and the x-axis is the number of high ratings (for our datasets, we consider 5 as a high rating). We can observe a rough trend: the larger the norm is, the more high ratings the corresponding item has, which means it is more popular. 
According to this intuition, the Max-Norm method selects the items with top-$M$ largest norms as an heuristic approximation of $\mathcal{Y}^*$, as described in Algorithm \ref{alg:max-norm}.
\begin{algorithm}
	\caption{Max-Norm}
	\label{alg:max-norm}
	\begin{algorithmic}[1]
		\STATE {\bfseries Input:} item embeddings $\mathcal{X}$
		\STATE Initialize $\mathcal{Y}=\emptyset$.
		\FOR {every item $x$ in $\mathcal{X}$}
		\STATE  Calculate the norm of $x$.
		\ENDFOR
		\STATE Sort the norms in descending order.
		\STATE Add the top-$M$ elements into. $\mathcal{Y}$
		\RETURN subset $\mathcal{Y}$ of size $M$
	\end{algorithmic}
\end{algorithm} 

\medskip

\begin{figure}
\centering
\subfloat[MovieLens-20m]{\label{fig:pop_20m}{\includegraphics[width=0.33\textwidth]{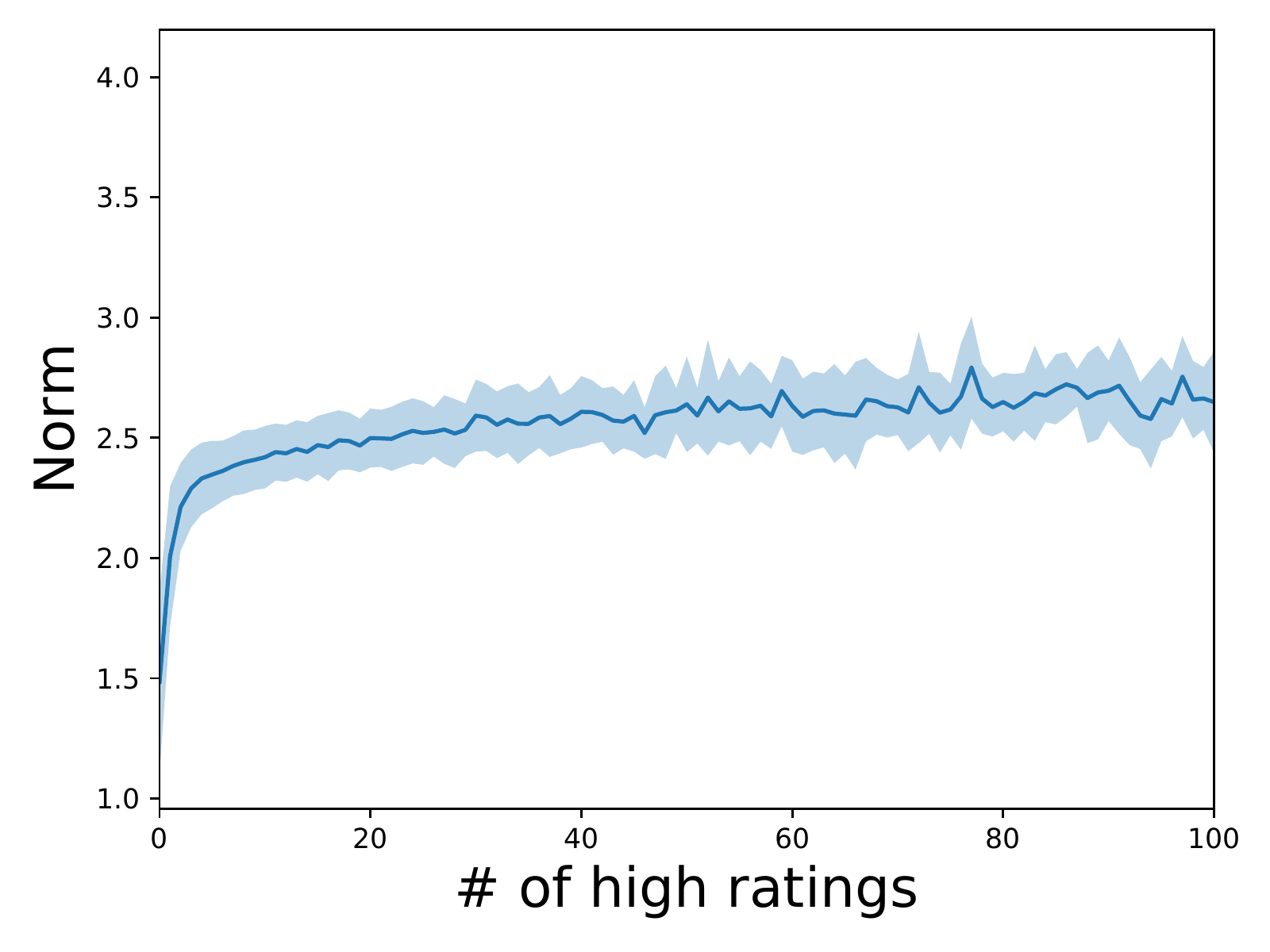}}}\hfill
\subfloat[Epinions]{\label{fig:pop_epin}{\includegraphics[width=0.33\textwidth]{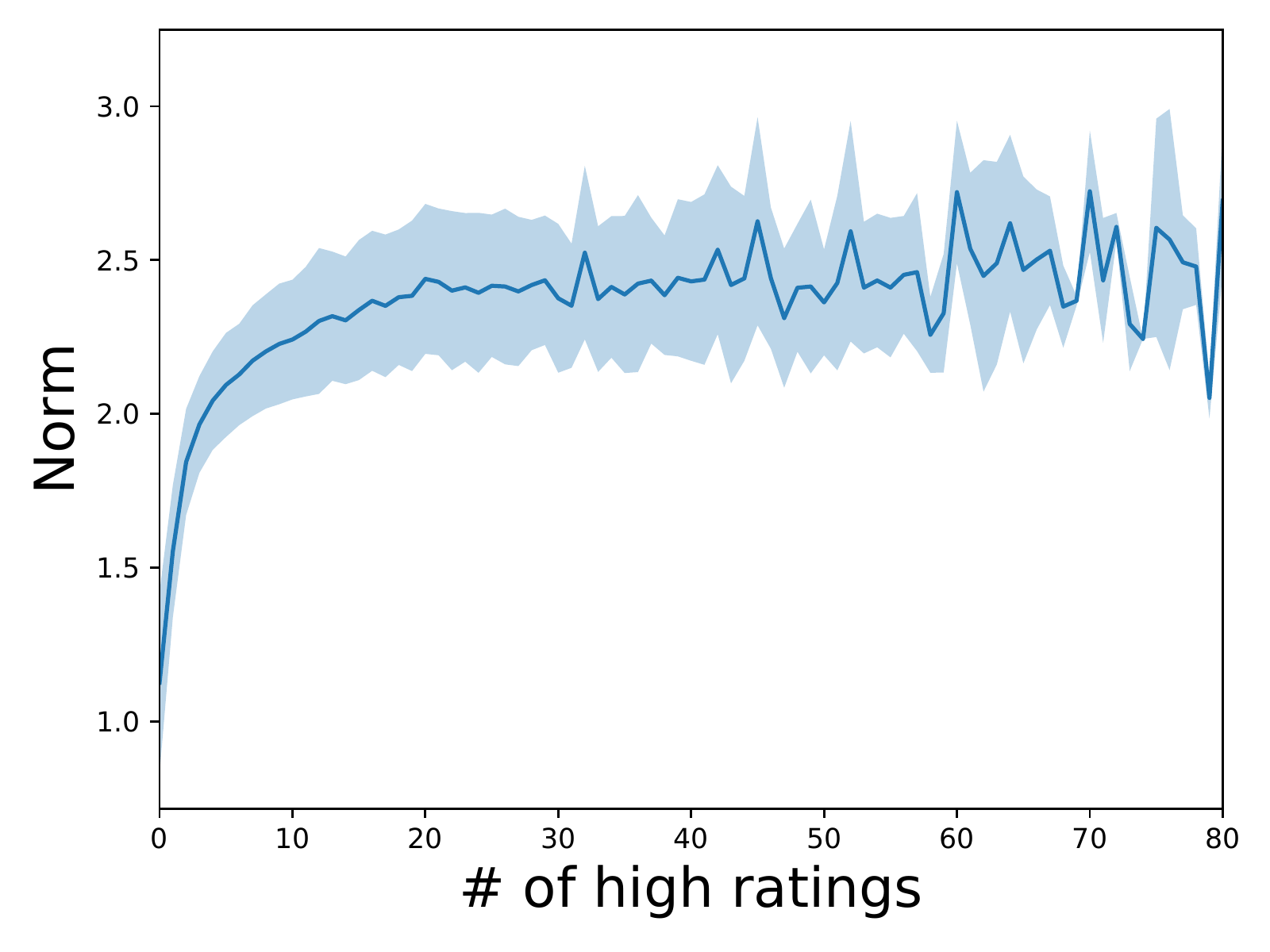}}}\hfill
\subfloat[Amazon's All Beauty]{\label{fig:pop_beauty}{\includegraphics[width=0.33\textwidth]{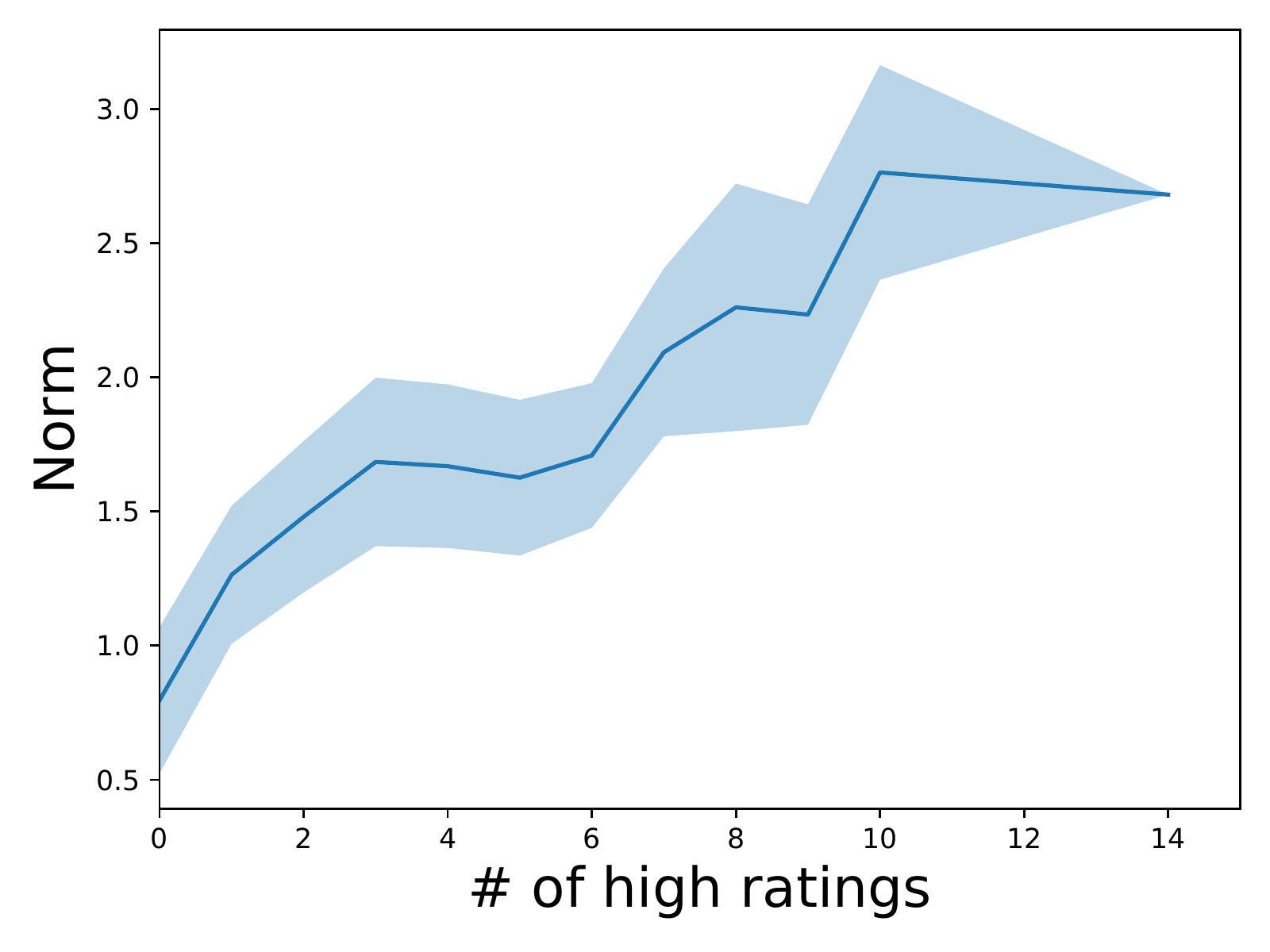}}}
\caption{The relationship between norm and the number of high ratings. For each value on the x-axis, we calculate the mean and variance of the norms of the corresponding items, and plot them as a line and shadow respectively.}
\label{fig:pop_norm}
\end{figure}


\medskip

\subsection{Max-In-Degree method}

\medskip
We introduce another heuristic method for item selection problem: we build inner product proximity graph using item latent vectors, then select the top-$M$ items with largest in-degree as the approximation to $\mathcal{Y}^*$. Inner product proximity graph is built by connecting each item (source) to his top-$K$ nearest neighbours (destinations) by a directed link. 
\begin{figure}
\centering
\includegraphics[width=0.45\textwidth]{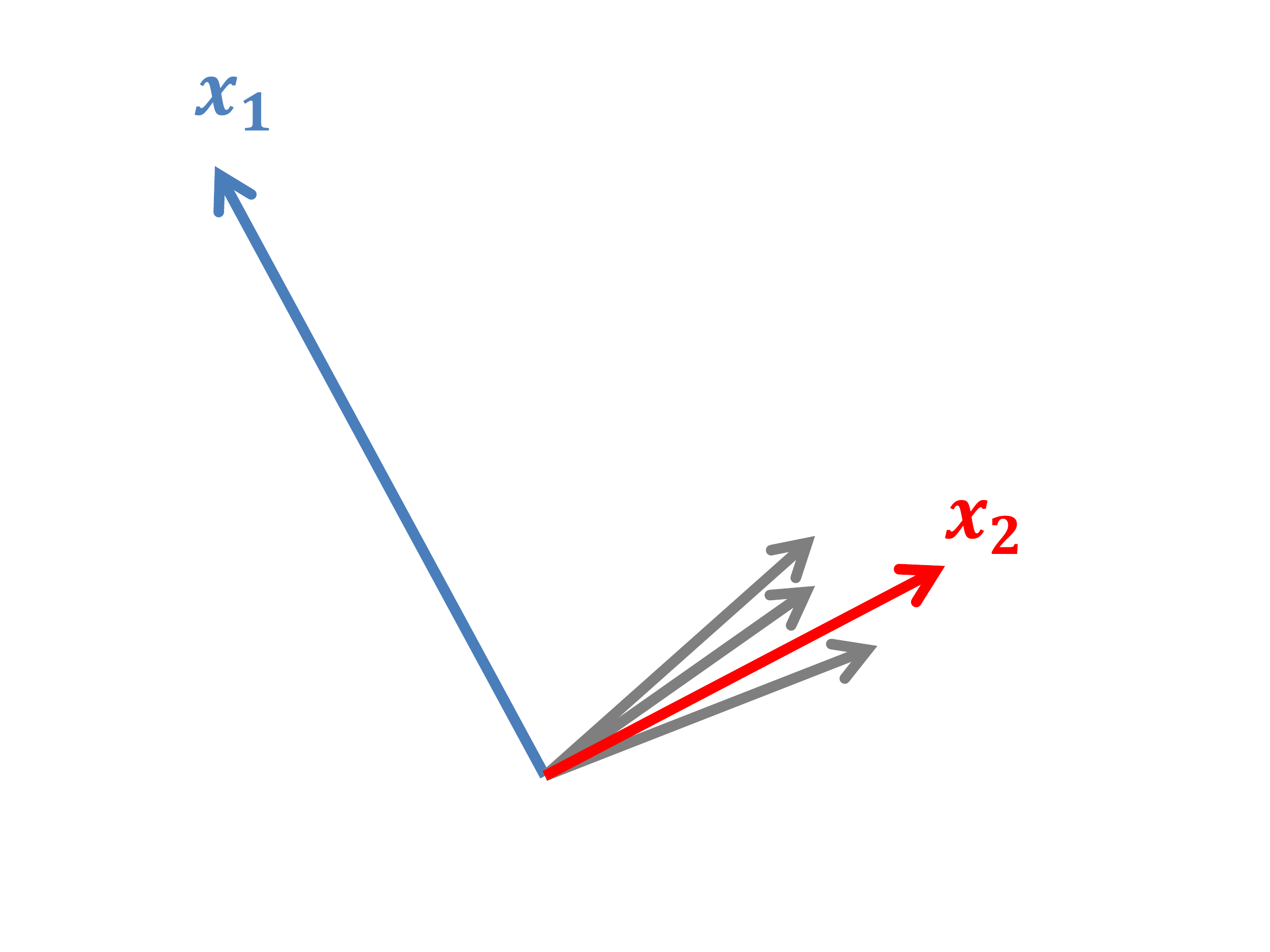}
\caption{A toy example of item vectors. There are totally five items. Max-Norm will select $x_1$ while Max-In-Degree will select $x_2$.}
\label{fig:maxnorm_indegree}
\end{figure}
If an item has a large in-degree, it means there are so many other items similar to this item. The physical meaning is that this item could be a potential substitution for many other items, and as a result, it gets more chance to be liked and consumed by users.
 Max-In-Degree favors the item with a large norm and in a direction where many other items concentrate.
 In Figure \ref{fig:maxnorm_indegree}, we use a toy example to illustrate the difference between the Max-Norm and the Max-In-Degree methods.
 Assuming there are totally five items, Max-Norm will recommend item $x_1$ because it has the largest norm. Max-In-Degree will recommend item $x_2$, because there are many other items in almost the same direction and $x_2$ has the largest norm among them.
 Max-Norm method is more likely to succeed on datasets with larger norm-bias, while Max-In-Degree tend to succeed on datasets where user and item latent vectors have a similar distribution.

However, building the exact inner product proximity graph requires a time complexity of $O(N^2)$. To accelerate, we use the ip-NSW~\cite{morozov:graphmips} algorithm to approximate it.
By using ip-NSW, our Max-In-Degree algorithm is described in Algorithm \ref{alg:max-in-degree}. The time complexity of Algorithm \ref{alg:max-in-degree} is $O(N(\log N))$.
\begin{algorithm}
	\caption{Max-In-Degree method}
	\label{alg:max-in-degree}
	\begin{algorithmic}[1]
		\STATE {\bfseries Input:} item embeddings $\mathcal{X}$
		\STATE Initialize $\mathcal{Y}=\emptyset$.
		\STATE  Construct the inner product proximity graph using Algorithm HNSW.
		\FOR {every item $x$ in $\mathcal{X}$}
		\STATE  Calculate the in-degree of $x$ in the graph.
		\ENDFOR
		\STATE Sort the in-degrees in descending order.
		\STATE Add elements corresponding to top-$M$ in-degrees into $\mathcal{Y}$.
		\RETURN the subset $\mathcal{Y}$ of size $M$
	\end{algorithmic}
\end{algorithm}

\subsection{User-Expectation method}
Since cold-start users come from the same distribution as the warm users, we can estimate a cold-start user's latent vector $q$ by the expectation of warm users, i.e.
\begin{align}
    q=\frac{1}{\vert \mathcal{U}\vert}\sum_{u\in \mathcal{U}}u.
\end{align}
Then the top-$M$ items with the largest $q^\mathsf{T}x$ value forms the approximation of $\mathcal{Y}^*$, as described in Algorithm \ref{alg:user-expectation}.
Since taking expectation loses information, the method may fail the case where the variance of user distribution is large.
\begin{algorithm}
	\caption{User-Expectation}
	\label{alg:user-expectation}
	\begin{algorithmic}[1]
		\STATE {\bfseries Input:} user embeddings $\mathcal{U}$, item embeddings $\mathcal{X}$
		\STATE Initialize $\mathcal{Y}=\emptyset$
		\STATE calculate    $q=\frac{1}{\vert \mathcal{U}\vert}\sum_{u\in \mathcal{U}}u.$
		\FOR {every item $x$ in $\mathcal{X}$}
		\STATE  Calculate the value of $q^{\mathsf{T}}x$.
		\ENDFOR
		\STATE Add the top-$M$ $x\in\mathcal{X}$ with largest $q^{\mathsf{T}}x$ values into $\mathcal{Y}$.
		\RETURN subset $\mathcal{Y}$ of size $M$.
	\end{algorithmic}
\end{algorithm} 

\subsection{Inner Product Graph Search (IPGS) method}
The main idea of the IPGS method is to statistic the favorite items for each warm user, then recommend the top-$M$ items with the highest frequency.
However, the complexity of doing so is $O(WN)$.
To reduce the high time complexity, we use a graph searching method to approximate the result. The algorithm for IPGS is described in Algorithm \ref{alg:IPGS}.
\begin{algorithm}
	\caption{IPGS}
	\label{alg:IPGS}
	\begin{algorithmic}[1]
		\STATE {\bfseries Input:} item latent vectors $X$, user latent vectors $U$
		\STATE Initialize $\mathcal{Y}=\emptyset$.
		\STATE  Construct the inner product proximity graph $\mathcal{G}$ for $\mathcal{X}$ using Algorithm~\ref{alg:max-in-degree}.
		\FOR {every query $u$ in $\mathcal{U}$}
		\STATE  Process query $u$ using HNSW query and find an item which is the top-1 inner product neighbor for $u$ in $\mathcal{G}$.
		\ENDFOR
		\STATE Count the frequency of each item $x$ in $\mathcal{X}$ being nominated as top-1 neighbor.
		\STATE Sort the frequencies in descending order.
		\STATE Add elements corresponding to top-$M$ frequencies into $\mathcal{Y}$.
		\RETURN the subset $\mathcal{Y}$ of size $M$
	\end{algorithmic}
\end{algorithm}

In Algorithm \ref{alg:IPGS}, we still construct exactly the same inner product proximity graph as Max-In-Degree method does. Instead of choosing the vertices with highest in-degree, we make use of user latent vectors as queries to search in the inner product graph. The search process is similar to the vertex insertion process during the construction, except that we do not need to connect the queries to approximate top-K neighbors in the end. In our implementation, we find the top-1 item for each existing users and count the frequency of each item being selected. The items with top-$M$ frequency after querying process is the our approximation to $\mathcal{Y}^*$. 
 The querying process can be seen as a part of the graph construction process yet without actual insertions. Querying process is computationally efficient, so it does not bring a significant burden to the computation. The time complexity of IPGS is $O((W+N)\log N)$
\subsection{Submodular greedy algorithm}

\medskip

Different from the previous heuristic methods, submodular greedy algorithm directly optimize the problem stated in Eq. \ref{eq:item_selection}. Eq. \ref{eq:item_selection} is a submodular set function, which allows us to apply the existing algorithms for submodular functions on it. 

Below we introduce some background knowledge of submodular functions before demonstrating the algorithm we actually used.

\subsubsection{Submodular Functions}
Submodular functions have the intuitive diminishing returns property. Formally, a submodular function $f$ assigns a subset $\mathcal{Y} \subseteq \mathcal{X}$ a utility value $f(\mathcal{Y})$ such that
\[
f(\mathcal{Y} \cup\{i\})-f(\mathcal{Y}) \geq f(\mathcal{Z} \cup\{i\})-f(\mathcal{Z})
\]
for any $\mathcal{Y} \subseteq \mathcal{Z} \subseteq \mathcal{X}$ and $i \in \mathcal{X} \backslash \mathcal{Z}$. We call $\mathcal{X}$ the ground set.\\

Note that this definition just means that adding an element $i$ to a subset $\mathcal{Y}$ of set $\mathcal{Z}$ yields at least much value (or more) as if we add $i$ to $\mathcal{Z}$. In other words, the marginal gain of adding $i$ to $\mathcal{Y}$ is greater or equal to the marginal gain of adding $i$ to $\mathcal{Z}$. The notation for the marginal gain is:
\[
\Delta(i | \mathcal{Y})=f(\mathcal{Y} \cup\{i\})-f(\mathcal{Y}).
\]

In Eq. \ref{eq:favloss}, considering the fact that the entire item set $\mathcal{X}$ is fixed and thus the upper bound $\sum_{u\in \mathcal{U}} \max_{x \in \mathcal{X}} u^\mathsf{T}x$ is also fixed. Thus, Eq. \ref{eq:item_selection} can be reduced to:
\begin{equation} \label{eq:def2}
\begin{split}
\mathcal{Y}^{*}=\operatorname{argmax}_{\mathcal{Y}:\{|\mathcal{Y}| = M, \mathcal{Y} \subseteq \mathcal{X}\}} \sum_{u\in\mathcal{U}} \max_{y \in \mathcal{Y} \subseteq \mathcal{X}} u^\mathsf{T}y
\end{split}
\end{equation}

\begin{proposition}
\label{prop:submodular}
$f(\mathcal{Y})=\sum_{u\in\mathcal{U}} \max_{y \in \mathcal{Y} \subseteq \mathcal{X}} u^\mathsf{T}y$ is a submodular function.
\begin{proof}
Given a subset $\mathcal{Y}$ and an extra element $z$, there are two possibilities: in the first case, $\max_{y \in \mathcal{Y}} u^\mathsf{T}y$ is equal to or greater than $u^\mathsf{T}z$, then $f(\mathcal{Y} \cup \{z\}) = f(\mathcal{Y})$; in the second case, $\max_{y \in \mathcal{Y}} u^\mathsf{T}y$ is less than $u^\mathsf{T}z$, then $f(\mathcal{Y} \cup \{z\}) > f(\mathcal{Y})$. Therefore, $f(\mathcal{Y} \cup \{z\}) \ge f(\mathcal{Y})$ always holds.
\end{proof}
\end{proposition}

We say that a submodular function is monotone if for any $\mathcal{Y} \subseteq \mathcal{Z} \subseteq \mathcal{X}$ we have $f(\mathcal{Y}) \le f(\mathcal{Z})$. Intuitively, this means that adding more elements to a set cannot decrease its value. Apparently, $f(\mathcal{Y})$ in Proposition \ref{prop:submodular} is monotonically increasing.

\subsubsection{Maximization of monotone submodular functions}
The problem of Eq. \ref{eq:def2} is NP-hard. Fortunately, by using submodularity and monotonicity, a simple greedy algorithm provides a solution with a nice approximation guarantee, which we will prove later. The algorithm starts with the empty set, and then repeats the following step for $i=0, \ldots,(M-1)$:

\[
\mathcal{Y}_{i+1}=\mathcal{Y}_{i} \cup\left\{\operatorname{argmax}_{x \in \mathcal{X} \backslash \mathcal{Y}_{i}} f\left(\mathcal{Y}_{i} \cup\{x\}\right)\right\}
\]

Note that 
\[
\left\{\operatorname{argmax}_{x \in \mathcal{X} \backslash \mathcal{Y}_{i}} f\left(\mathcal{Y}_{i} \cup\{x\}\right)\right\}=\left\{\operatorname{argmax}_{x \in \mathcal{X} \backslash \mathcal{Y}_{i}} \Delta\left(x | \mathcal{Y}_{i}\right)\right\}
\]

The pseudo code of the greedy algorithm is shown in Algorithm \ref{alg:submodular}. 
The runtime of Algorithm \ref{alg:submodular} is $O(WMN)$ (recall that $W=\vert \mathcal{U} \vert$, $M=\vert \mathcal{Y} \vert$ and $N=\vert \mathcal{X} \vert$). Because there are $O(MN)$ function evaluations and in each evaluation we have to compute $W$ inner products with $W$ users.
Next, we show the guarantee of the greedy algorithm and its proof in Theorem \ref{theorem:submodular}. 

\begin{algorithm}
	\caption{Submodular Greedy Algorithm}
	\label{alg:submodular}
	\begin{algorithmic}[1]
		\STATE {\bfseries Input:} item latent vectors $\mathcal{X}$, user latent vectors $\mathcal{U}$
		\STATE Initialize $\mathcal{Y}=\emptyset$ 
		\STATE Find the first greedy solution, $y_1 = \operatorname{argmax}_{x \in \mathcal{X}} \sum_{u \in \mathcal{U}}u^\mathsf{T}x$
		\STATE Add $y_1$ into $\mathcal{Y}$ and record current max inner product value for each $u \in \mathcal{U}$
		\WHILE{$|\mathcal{Y}|<M$}
		\FOR {every item $x$ in $\mathcal{X}/\mathcal{Y}$}
		\STATE  Calculate $f(\mathcal{Y} \cup x)=\sum_{u \in \mathcal{U}} \max_{v \in \{\mathcal{Y} \cup x\}} u^\mathsf{T}v$
		\ENDFOR
		\STATE Sort the results and add the item $x$ with the max $f(\mathcal{Y} \cup x)$ value into $\mathcal{Y}$
		\STATE Update the current max inner product value for each $u \in \mathcal{U}$
		\ENDWHILE
		\RETURN the subset $\mathcal{Y}$ of size $M$
	\end{algorithmic}
\end{algorithm} 

\medskip
\begin{theorem}\label{theorem:submodular}
Let $\mathcal{Y}_i = (x_1, x_2, \ldots, x_i)$ be the chain formed by the greedy algorithm and $\mathcal{Y}^{*} = (x_1^{*}, x_2^{*}, \ldots, x_M^{*})$ be the optimal solution in an arbitrary order. And let $OPT = f(\mathcal{Y}^{*})$ be the value of the optimal solution.

\medskip

We have 
\begin{equation} \label{eq:greedy}
\begin{split}
f\left(\mathcal{Y}_{M}\right) \geq(1-1 / e) OPT
\end{split}
\end{equation}

Note that $1-1 / e \approx 0.63$.

\bigskip

\begin{proof}
For all $i \le M$, we have:\\
\begin{equation*} \label{eq:greedy_1}
\begin{split}
f\left(\mathcal{Y}^{*}\right) &\leq f\left(\mathcal{Y}^{*} \cup \mathcal{Y}_{i}\right)   \qquad \qquad \qquad \textrm{Monotonicity}\\
&= f(\mathcal{Y}_i) + \sum_{j=1}^{M} \Delta\left(x_{j}^{*} | \mathcal{Y}_{i} \cup\left\{x_{1}^{*}, x_{2}^{*}, \ldots, x_{j-1}^{*}\right\}\right)\\
&\le f\left(\mathcal{Y}_{i}\right)+\sum_{z \in \mathcal{Y}^{*}} \Delta\left(z | \mathcal{Y}_{i}\right)  \qquad  \textrm{Using submodularity}\\
&\le f\left(\mathcal{Y}_{i}\right)+\sum_{z \in \mathcal{Y}^{*}} \Delta\left(x_{i+1} | \mathcal{Y}_{i}\right)  \qquad  x_{i+1} = \operatorname{argmax}_{x \in \mathcal{X} \backslash \mathcal{Y}_{i}} \Delta\left(x | \mathcal{Y}_{i}\right)\\
&= f\left(\mathcal{Y}_{i}\right) + M\Delta\left(x_{i+1} | \mathcal{Y}_{i}\right)
\end{split}
\end{equation*}

That is
\begin{equation} \label{eq:greedy_2}
\begin{split}
\Delta\left(x_{i+1} | \mathcal{Y}_{i}\right) \ge \frac{1}{M}(OPT - f\left(\mathcal{Y}_{i}\right))
\end{split}
\end{equation}

Now we define $\delta_{i} = OPT - f(\mathcal{Y}_{i})$, which implies that 
\[
\delta_{i} - \delta_{i+1} = f(\mathcal{Y}_{i+1}) - f(\mathcal{Y}_{i}) = \Delta(x_{i+1} | \mathcal{Y}_{i}).
\]

In other words, we have proved that the element added at iteration $i+1$ by the greedy algorithm reduces the gap to the optimal solution by a significant amount - by at least $\frac{1}{M}(\textrm{OPT} - f(\mathcal{Y}_i))$. Another way to write the same equation is
\begin{equation} \label{eq:greedy_3}
\begin{split}
\delta_{i+1} \leq\left(1-\frac{1}{M}\right) \delta_{i}
\end{split}
\end{equation}
If we recursively apply this definition, we have that
\begin{equation} \label{eq:greedy_4}
\begin{split}
\delta_{M} \leq\left(1-\frac{1}{M}\right)^{M} \delta_{0}
\end{split}
\end{equation}

Now, $\delta_{0}=O P T-f(\emptyset) \leq O P T$. Thus, using the well-known bound $1-x \leq e^{-x}$ for $x \in \mathbb{R}$, we have that
\begin{equation} \label{eq:greedy_5}
\begin{split}
\delta_{M} = OPT - f\left(\mathcal{Y}_{M}\right) \leq\left(1-\frac{1}{M}\right)^{M} O P T \leq \frac{1}{e} O P T
\end{split}
\end{equation}
Or equivalently:
\begin{equation} \label{eq:greedy_6}
\begin{split}
f\left(\mathcal{Y}_{M}\right) \geq(1-1 / e) OPT,
\end{split}
\end{equation}
which concludes the proof.

\end{proof}
\end{theorem}
\medskip

\subsection{Approximate convex hull}
In this section, we try to find the convex hull of $\mathcal{X}$, denoted as $Conv(\mathcal{X})$ and use it as recommendation.
As stated in Theorem \ref{theorem:convexhull}, we observe that $\mathcal{Y}^*$ in Eq. \ref{eq:item_selection} must be belong to the convex hull of $\mathcal{X}$. 

\begin{theorem}
\label{theorem:convexhull}
In Eq. \ref{eq:item_selection}, for any $\mathcal{U}$, $\mathcal{X}$ and $M$, $\mathcal{Y}^* \subset Conv(\mathcal{X})$.

\begin{proof}
We prove by contradiction.
If there exists a $y_i\in\mathcal{Y}$ that $y_i\not\in Conv(\mathcal{X})$, then we can find a substitute $x\in Conv(\mathcal{X})$ such that the $fav\_loss$ is smaller.
Suppose $Conv(\mathcal{X})=\{x_1,x_2,...,x_D\}$,  $y_i=\sum_{d=1}^D \alpha_d x_d$ is the convex combination of $y_i$ and we denote $Q(y_i):=\{q\in Q \vert y_i=\operatorname{argmax}_{y \in \mathcal{Y}}q^\mathsf{T}y\}$. Then, we have,
\begin{flalign}
   loss(\mathcal{Y}) &= \sum_{q\in Q}[\max_{x \in \mathcal{X}} q^\mathsf{T}x - \max_{y \in \mathcal{Y}} q^\mathsf{T}y]\\
   &=\sum_{q\in Q}\max_{x \in \mathcal{X}} q^\mathsf{T}x -\sum_{q\in Q} \max_{y \in \mathcal{Y}} q^\mathsf{T}y\\
   &=\sum_{q\in Q}\max_{x \in \mathcal{X}} q^\mathsf{T}x -[\sum_{q\in Q\setminus Q(y_i)} \max_{y \in \mathcal{Y}} q^\mathsf{T}y+\sum_{q\in Q(y_i)}  q^\mathsf{T}y_i]\\
   &=\sum_{q\in Q}\max_{x \in \mathcal{X}} q^\mathsf{T}x -[\sum_{q\in Q\setminus Q(y_i)} \max_{y \in \mathcal{Y}} q^\mathsf{T}y+\sum_{q\in Q(y_i)}  q^\mathsf{T}\sum_{d=1}^D \alpha_d x_d]\\
   &=\sum_{q\in Q}\max_{x \in \mathcal{X}} q^\mathsf{T}x -[\sum_{q\in Q\setminus Q(y_i)} \max_{y \in \mathcal{Y}} q^\mathsf{T}y+\sum_{d=1}^D \alpha_d (\sum_{q\in Q(y_i)}  q)^\mathsf{T}x_d]
\end{flalign}
(suppose $x_j=\operatorname{argmax}_{x \in Conv(\mathcal{X})}(\sum_{q\in Q(y_i)}  q)^\mathsf{T}x$)
\begin{flalign}
&>\sum_{q\in Q}\max_{x \in \mathcal{X}} q^\mathsf{T}x -[\sum_{q\in Q\setminus Q(y_i)} \max_{y \in \mathcal{Y}} q^\mathsf{T}y+  (\sum_{q\in Q(y_i)}  q)^\mathsf{T}x_j]\\
&\geq loss((\mathcal{Y} \setminus y_i)\cup x_j),
\end{flalign}
which conclude the proof.
\end{proof}
\end{theorem}

According to theorem \ref{theorem:convexhull}, there is a corollary,
\begin{corollary}
    If $\vert Conv(X) \vert  \leq M$, then
    \begin{align}
         fav\_loss(Conv(\mathcal{X}))=\min_{\mathcal{Y}:\{\vert \mathcal{Y} \vert=M, \mathcal{Y}\subset\mathcal{X}\}}fav\_loss(\mathcal{Y}).   
    \end{align}

\end{corollary}

Ideally, being given the convex hull can help reduce the computation costs. If $\vert Conv(X) \vert > M$, we can use this subset $Conv(X)$ for further selection instead of the entire set $\mathcal{X}$. And if $|Conv(X)| \le M$, it is already the optimal solution. 

\medskip


Unfortunately, the cardinality of $ Conv(\mathcal{X}) $ tends to greatly increase in high dimension. In our problem setting, the size of recommendation subset is at most 200 while the dataset cardinality is from 26k to 270k. The size of $Conv(\mathcal{X})$ is much larger than 200, sometimes even makes up more than a third of the entire $ \mathcal{X}$. Therefore, this method does not help much in terms of efficiency and accuracy. Also, finding the convex hull is computationally expensive~\cite{sartipizadeh:computing}. We will not show its result in the experiment section since other methods outperform it significantly.



\section{Experiments}
\subsection{Datasets}
\label{sec:usercold_datasets}
We evaluate our proposed methods and various baselines on three public recommender datasets: MovieLens-20m~\cite{ml_20m}, Epinions~\cite{epinions} and Amazon's All Beauty~\cite{beauty}. The data statistics are summarised as follows:
\begin{table}[h]
	\centering
	\caption{Dataset statistics}
	\label{tab:datasets}

			\begin{tabular}{|c|c|c|c|}
\hline
                   & MovieLens-20M & Epinions & All Beausty \\ \hline
\#items            & 25,343        & 123,296  & 266,239     \\ \hline
\#warm users       & 110,795       & 32,131   & 26,069      \\ \hline
\#cold-start users & 27,698        & 8,032    & 6,517       \\ \hline
\end{tabular}

\end{table}
\subsection{Evaluation methodology}
We split the users in each dataset into warm users (trainset) and cold-start users (testset) with ratio 4:1.
The latent vectors of warm users and all items are derived by applying probabilistic matrix factorization~\cite{PMF} on all the ratings of warm users. Probabilistic matrix factorization is a powerful latent factor model that is simple and effective for recommender systems. 
These latent vectors are used as input of all baselines.
The selected items of various baselines are recommended to all cold-start users.
We first evaluate the $fav\_loss$ of various baselines on the warm users, to see which method can best solve Eq \ref{eq:item_selection}.
Then we evaluate this loss on the cold-user, to see if the selected items from warm users are still the favorites of cold start users.
Computing the $fav\_loss$ on cold-start users requires the their latent vectors. We derive the latent vectors $\mathcal{Q}=\{q_s\in \mathbb{R}^D\}_{s=1}^S$ of cold start users by factorizing the rating matrix of cold-start users with the fixed item vectors derived from warm users.

Apart from the $fav\_loss$, we are also interested in how our recommendation perform under other evaluation metric.
We consider three widely-used evaluation metrics for recommender systems:
Precision, Mean Average Precision (MAP) and Normalized Discounted Cumulative Gain (NDCG).
Their definitions are as follows:

\textbf{Precision}.
\begin{align}
Precision@M(u,\mathcal{Y}):=\frac{\sum_{r=1}^{M}\mathds{1}[\mathcal{Y}(r)\in Top(u,M)]}{M}.
\end{align}
where $\mathcal{Y}$ is a recommendation list with a size of at least $M$ and $Top(u,M)$ is the set of top-$M$ items with the largest inner products with user $u$.

\textbf{Average Precision}.
\begin{align}
AP@M(u,\mathcal{Y}):=\frac{\sum_{r=1}^{M} \mathds{1}[\mathcal{Y}(r) \in Top(u,M)] \times Precision@r(u,\mathcal{Y})}{M}.
\end{align}
$MAP@M$ is the mean of $AP@M$ across all users.

\textbf{NDCG}.
\begin{align}
NDCG@M(u,\mathcal{Y}):=Z_M \sum_{r=1}^M \frac{2^{\mathds{1}[\mathcal{Y}(r)\in Top(u,M)]-1 }}{\log (r +1)},
\end{align}
where $Z_M$ is chosen
such that the perfect ranking has an NDCG@M value of 1.
\medskip

For all the three metrics, we report the average value over all users.
\subsection{Evaluation Results}
\subsubsection{$fav\_loss$ on warm and cold-start users}
\begin{figure}
\centering
\includegraphics[width=1.0\textwidth]{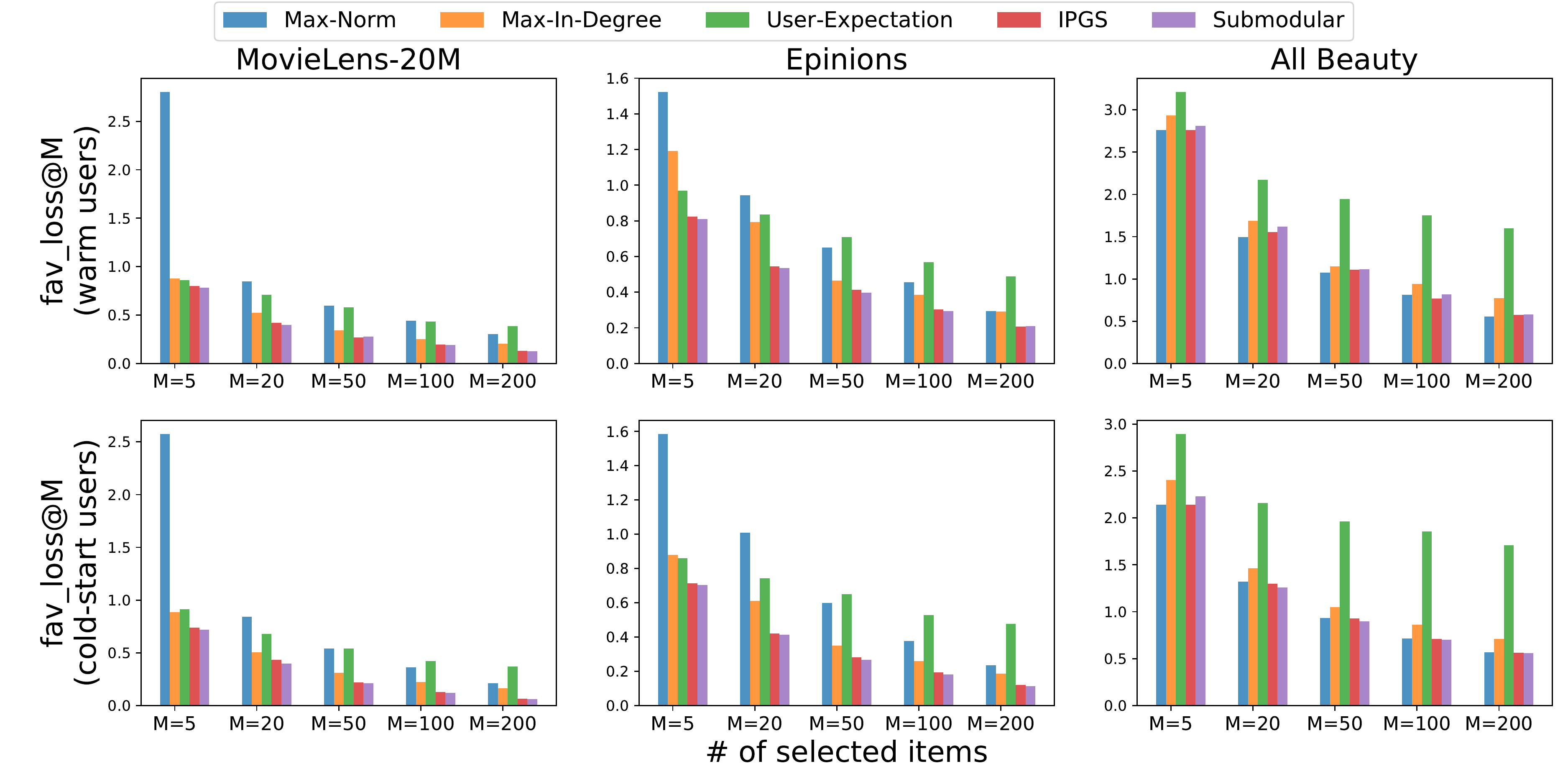}
\caption{fav\_loss on warm users and cold-start users.}
\label{fig:favloss_results}
\end{figure}
The $fav\_loss$ on warm users on the three datasets is shown in the first row of Figure \ref{fig:favloss_results}. 
The Submodular and IPGS methods perform best among all baselines most of the time. The good performance of Submodular is probably due to it is a principled method that directly approximates the $fav\_loss$.
The simple heuristic IPGS is also a good estimation to Eq.\ref{eq:item_selection}, probably because computing the favorite items of each user already filters the most useful information.
On the contrary, the User-Expectation method performs much worse than them and sometimes even got the worst performance. This indicates that the warm user vectors may be spread in several directions and the expectation is a coarse estimation. The performance of Max-In-Degree is only slightly inferior to that of IPGS, although it does not use user latent vectors. This implies that user latent vectors may have a similar distribution as the item latent vectors.

\begin{figure}[ht]
\centering
\subfloat[MovieLens-20m]{\label{fig:norm_20m}{\includegraphics[width=0.33\textwidth]{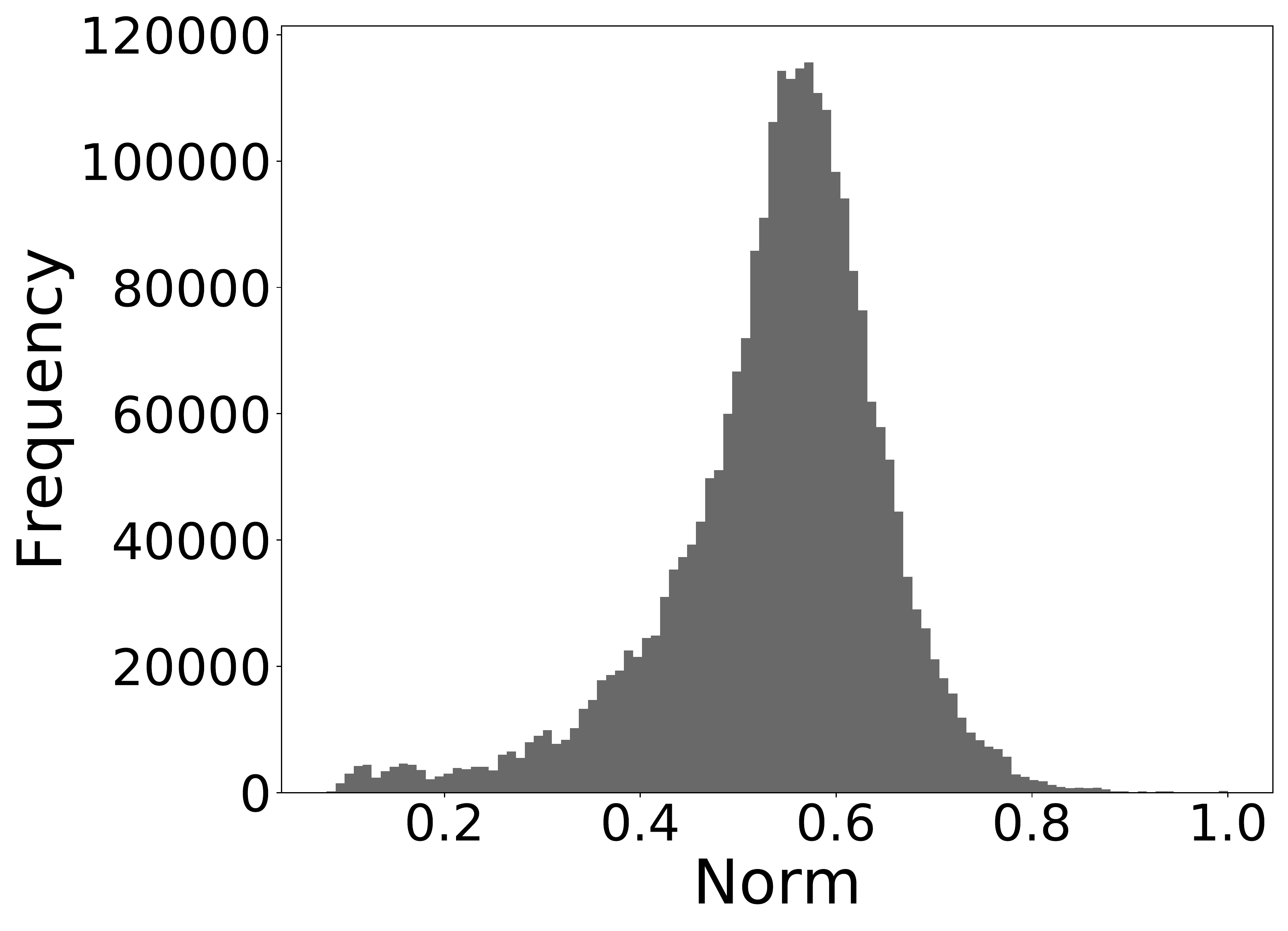}}}\hfill
\subfloat[Epinions]{\label{fig:norm_epin}{\includegraphics[width=0.33\textwidth]{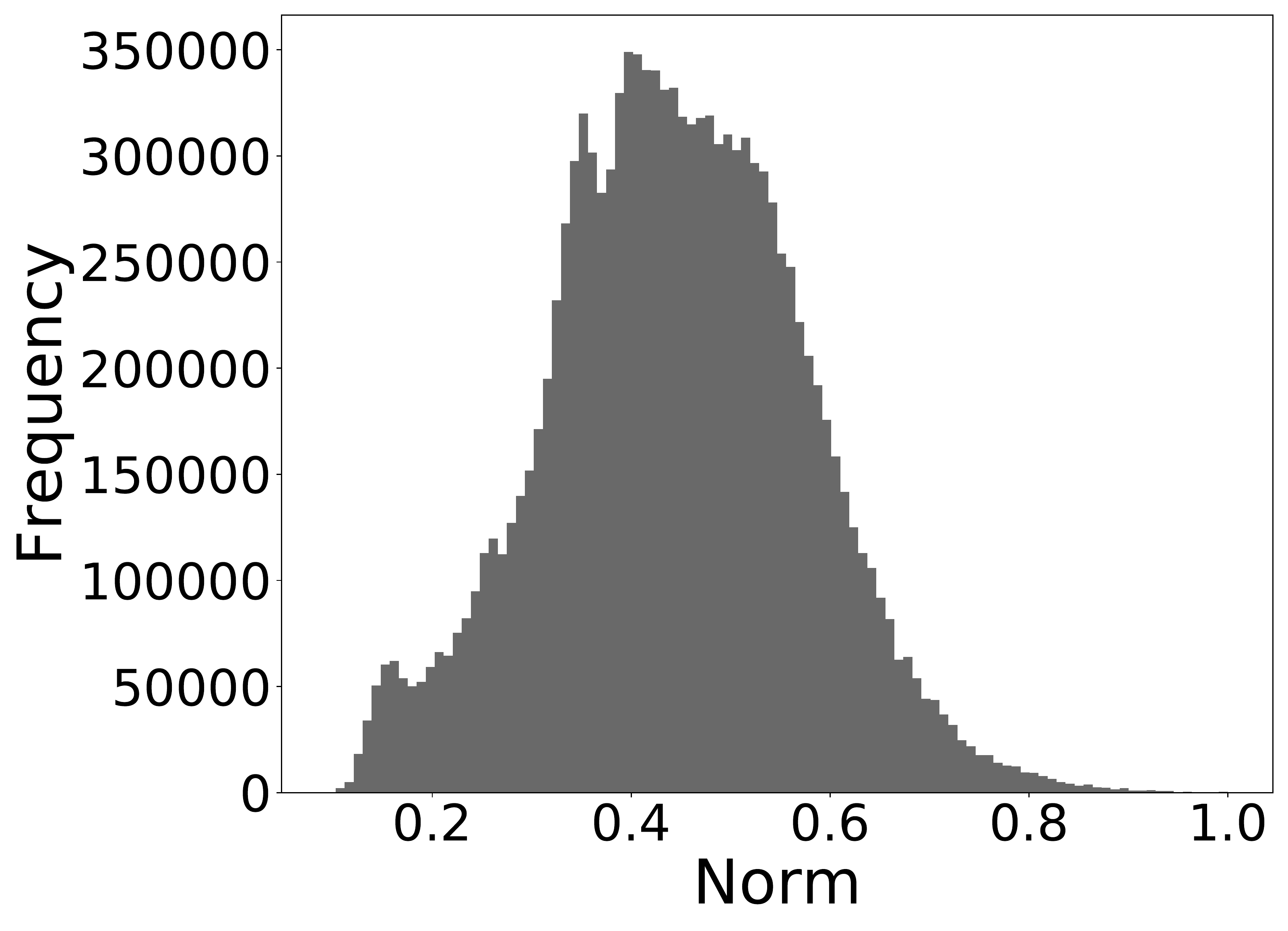}}}\hfill
\subfloat[Amazon's All Beauty]{\label{fig:norm_beauty}{\includegraphics[width=0.33\textwidth]{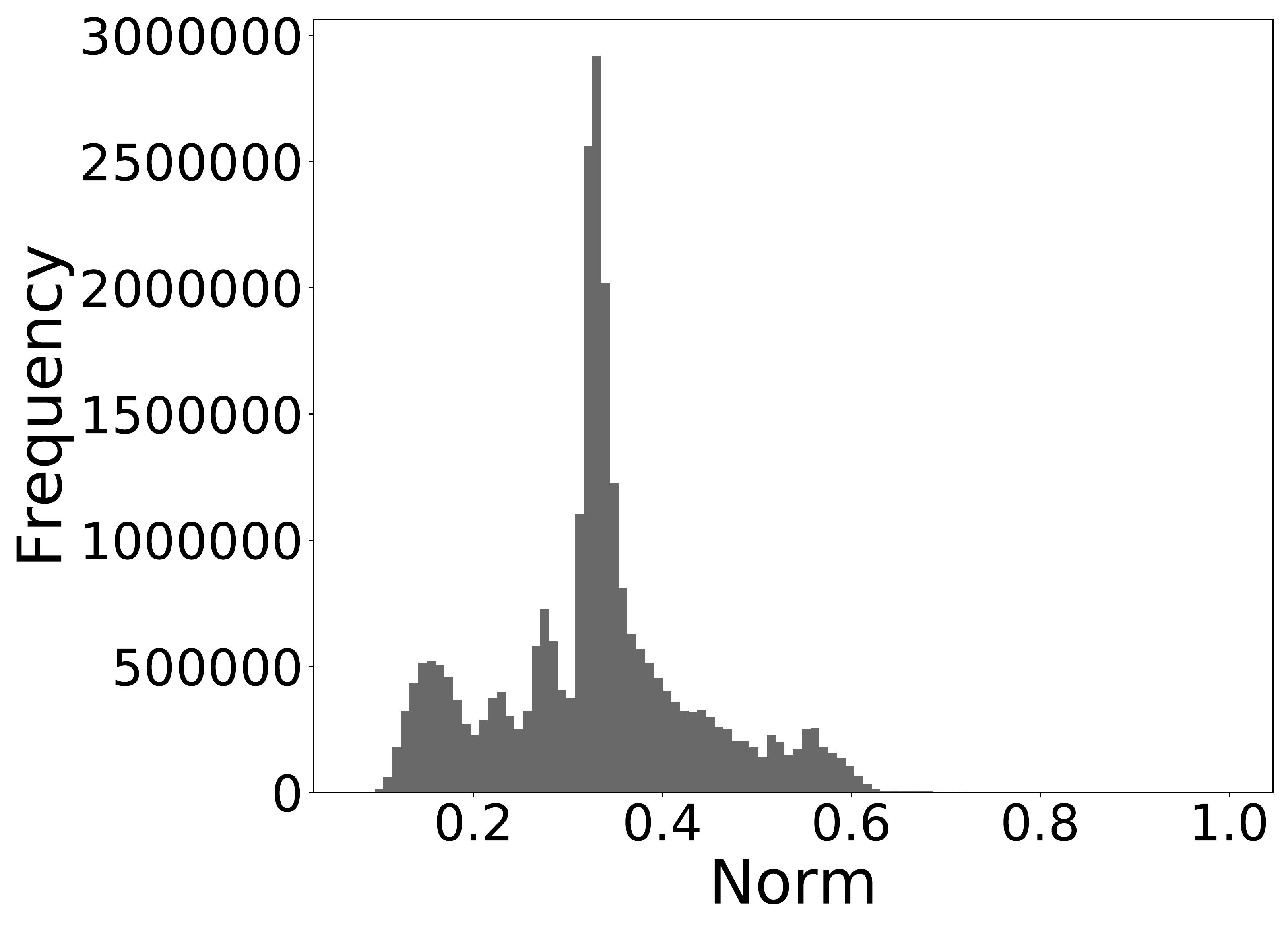}}}
\caption{Norm distributions of three datasets.}
\label{fig:norm_distribution}
\end{figure}

\begin{figure}[ht]
\centering
\subfloat[MovieLens-20m]{\label{fig:topk_20m}{\includegraphics[width=0.33\textwidth]{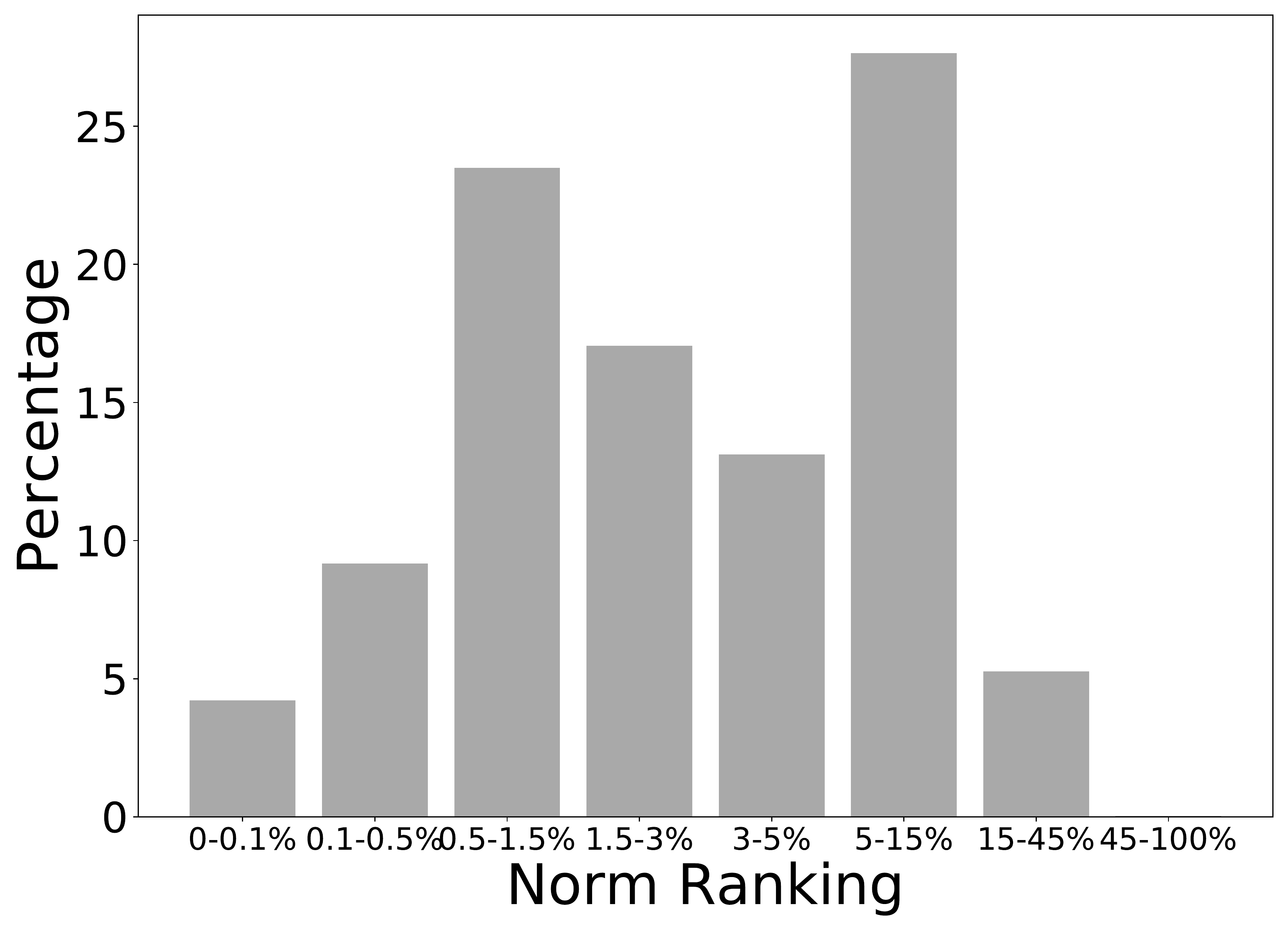}}}\hfill
\subfloat[Epinions]{\label{fig:topk_epin}{\includegraphics[width=0.33\textwidth]{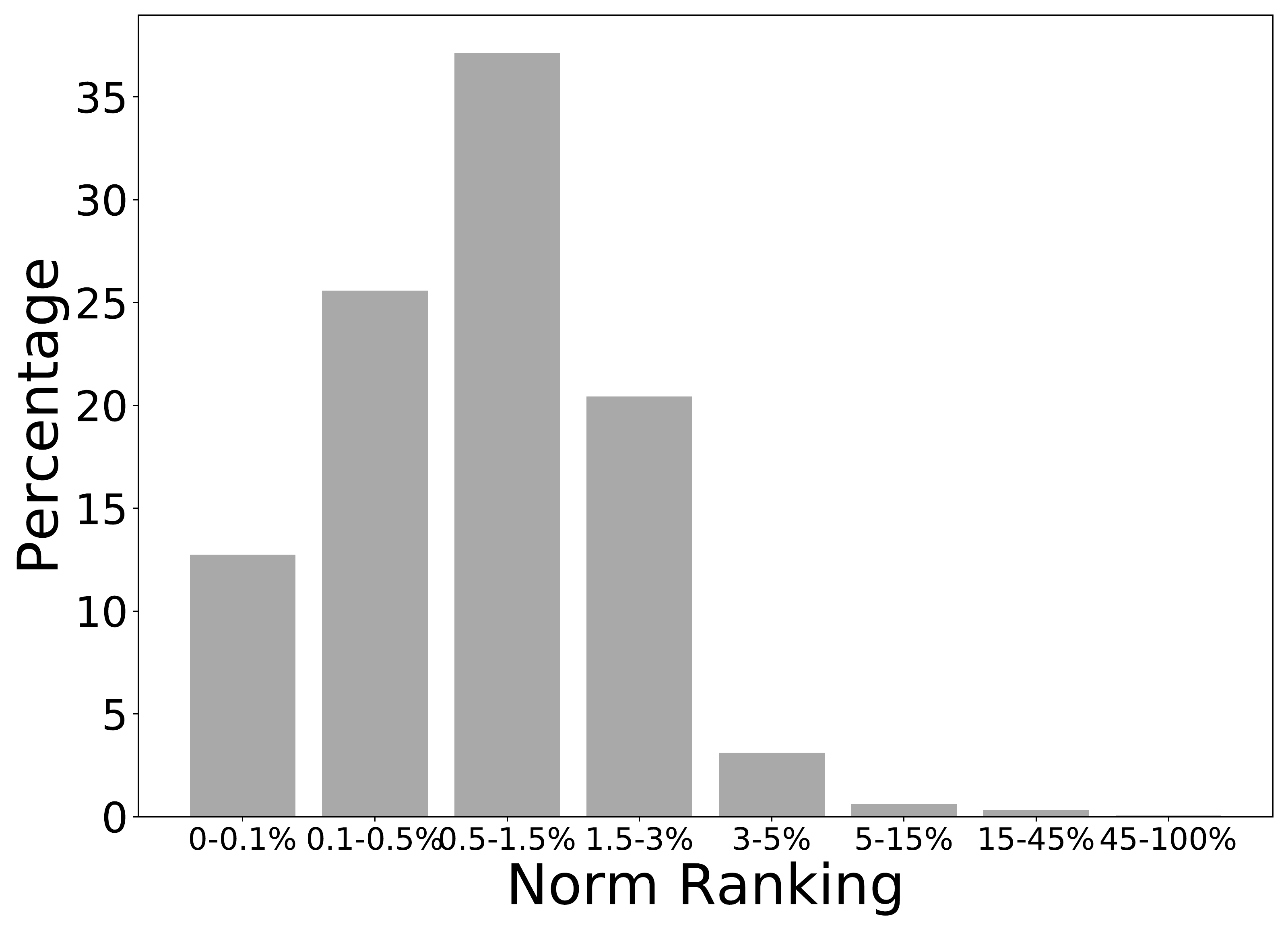}}}\hfill
\subfloat[Amazon's All Beauty]{\label{fig:topk_beauty}{\includegraphics[width=0.33\textwidth]{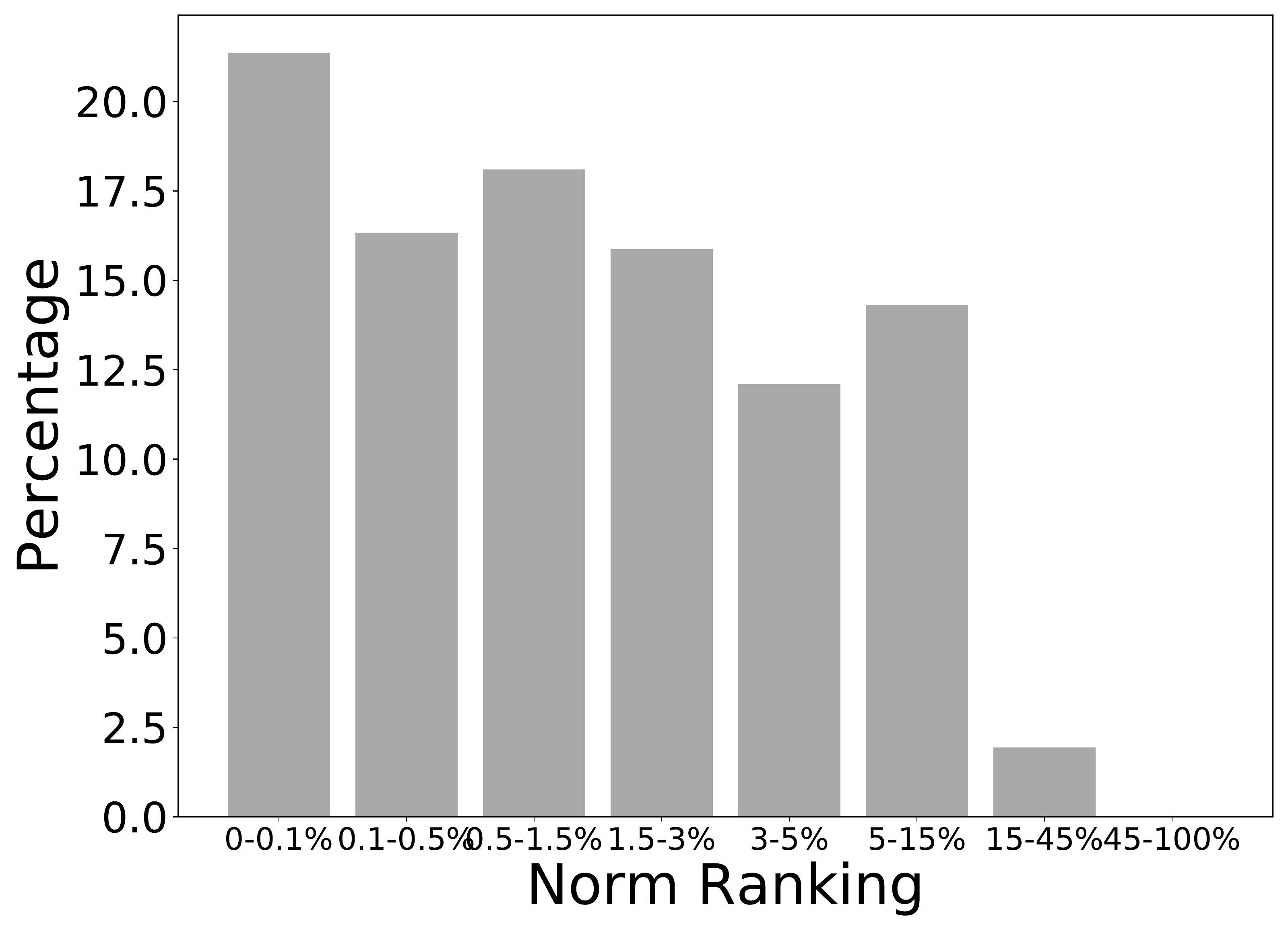}}}
\caption{The percentage that items in each norm group occupy in the users' top-10 favorite items.}
\label{fig:topk_norm_ranking}
\end{figure}

The Max-Norm method is one of the worst baselines on the MovieLens and Epinions datasets, but performs best on the $All\_Beauty$ dataset. 
In order to figure out the reason behind, we plot two figures to offer more statistics about the datasets. The first figure Fig.\ref{fig:norm_distribution} shows the norm distribution of all three datasets, where the norm of each vector is divided by the max norm of the dataset (so the max value on the x-axis is 1.0). For the second figure, we find the exact top-10 MIPS result of each user using a linear scan for each dataset, which gave us a result set containing 10*$|\mathcal{U}|$ items (duplicate items exist since an item can be in the results of multiple queries). Then we partition \textit{all} the items into groups according to their norms, e.g., items ranking top 0.1\% in the norm. Finally, for items in each norm group, we calculated the percentage they occupy in the formerly computed result set, which is plotted in Fig. \ref{fig:topk_norm_ranking}.

\medskip

From Fig.\ref{fig:norm_distribution}, we can observe that the norm difference in All Beauty is more significant than that in MovieLens-20m and Epinions. The medians of MovieLens-20m, Epinions and All Beauty are 0.669, 0.487 and 0.340, respectively. In All Beauty dataset, the norm of a large portion of items is only about 34\% of the max norm, which means the max-norm items are much longer than the rest items in terms of the norm. Since the norm-bias is more extreme,  the large-norm items are more likely to appear in the MIPS result set. 
This is supported by Fig.\ref{fig:topk_norm_ranking}, where we find that items with top 0.1\% norm occupy more than 20\% of the result set for All Beauty, while this figure on MovieLens-20m and Epinions are only 4.22\% and 12.75\% respectively. It is probably the larger norm-bias on All Beauty that makes Max-Norm more successful on it than on MovieLens-20m and Epinions.

The data patterns are almost the same across different values of $M$, on the three datasets respectively. The $fav\_loss$ decreases as $M$ increases because the more items selected the more users could be satisfied. We also note that the Max-Norm method is more sensitive to the $M$ value on MovieLens and Epinions. When $M$ is small, like $M=5$ or $20$, the gap between Max-Norm and other baselines is much larger than that when $M$ is relatively large. This may be because the norm-bias on MovieLens and Epinions is not large enough, and thus the first few items with largest norms may not represent the optimal solution well.

The $fav\_loss$ of vaious baselines on cold-start users shows a very similar pattern as on the warm users. This fits our expectation because we assume that the cold-start users come from the same distribution as the warm users.

\subsubsection{Other metrics on cold-start users}
\begin{figure}[t]
\centering
\includegraphics[width=1.0\textwidth]{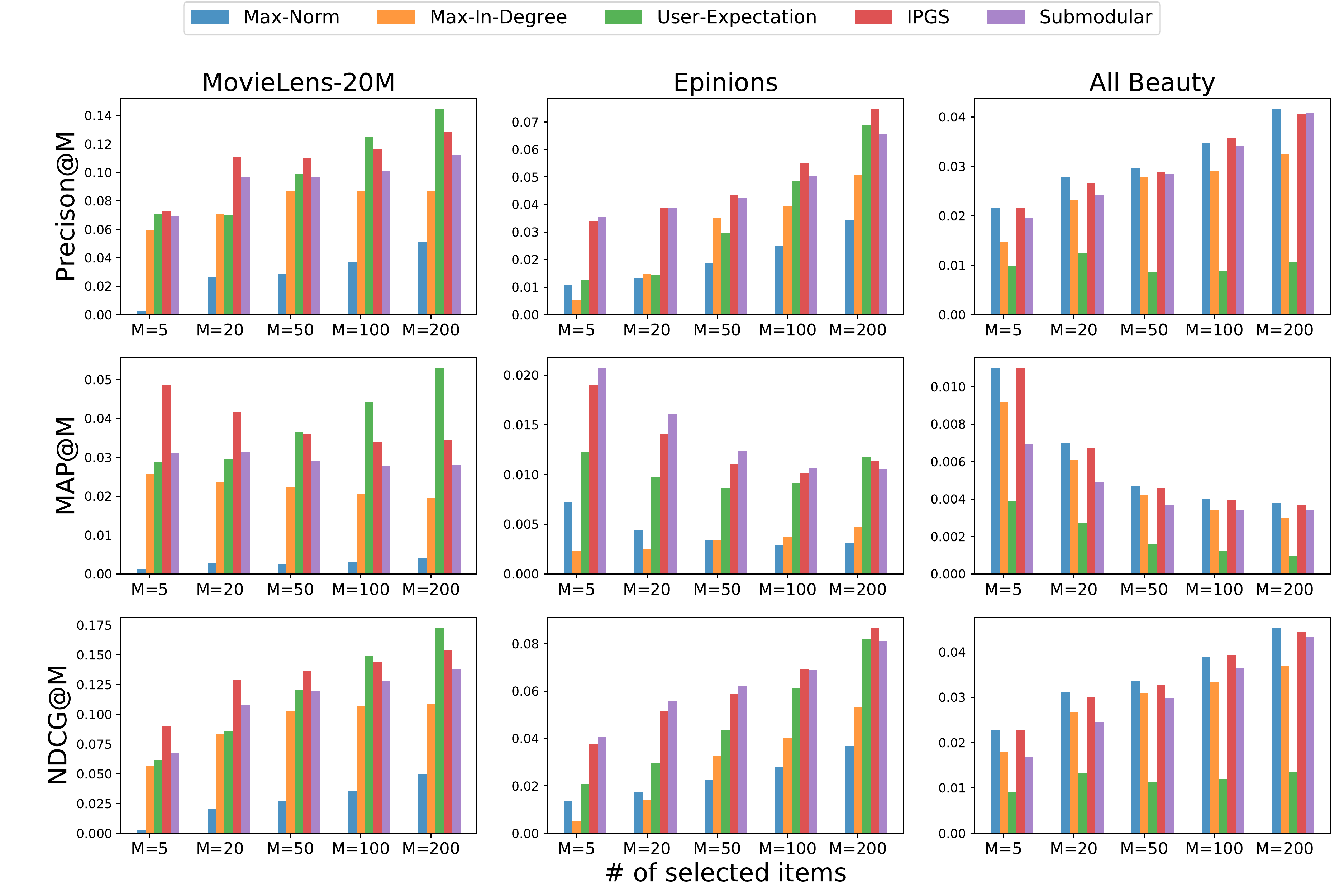}
\caption{Precision, MAP and NDCG on cold-start users.}
\label{fig:rec_results}
\end{figure}

However, as shown in Fig. \ref{fig:rec_results}, the baselines may perform in a different way on the other three metrics, i.e. Precision, MAP and NDCG, because these metrics are not directly related to our optimization objective.
On MovieLens-20M, the IPGS method acheives best performance on the three metrics when $M$ is small, while User-Expectation gradually outperforms the others when $M$ increases. On Epinions, IPGS and Submodular are still the best performing baselines. On the All Beauty dataset, Max-Norm and IPGS get the best performance, which is closely followed by the Submodular and Max-In-Degree methods.

\bigskip


\section{Conclusion}

In this chapter, we consider the pure user cold-start scenario, where neither interactions nor side information is available, and no user effort is required. We formulate the problem as an extension of the traditional MIPS problem, to which we propose six potential solutions. We conducted extensive experiments to evaluate the proposed solutions. Experimental results show that for the $fav\_loss$, it is desirable to use the IPGS method because it achieves a very low loss value and also has a relatively low computational complexity. Besides, IPGS also achieves the best performance on the other three recommendation metrics most of the time. Although submodular performs similar to IPGS, it is more computationally expensive.
For datasets with very large norm-bias, one can also consider using the Max-Norm method which even has a lower time complexity than IPGS.
%
%
%
%
\bibliographystyle{abbrv}
\bibliography{ref}
\end{document}